\documentclass[conference]{IEEEtran}
\IEEEoverridecommandlockouts
\usepackage{xcolor,soul,framed} 

  \IEEEoverridecommandlockouts
    \usepackage{cite}
    \usepackage{amsmath,amssymb,amsfonts}
    \usepackage{algorithmic}
    \usepackage{graphicx}
    \usepackage{textcomp}
    \usepackage{xcolor}
    \def\BibTeX{{\rm B\kern-.05em{\sc i\kern-.025em b}\kern-.08em
    T\kern-.1667em\lower.7ex\hbox{E}\kern-.125emX}}
    
\usepackage{cite}
\usepackage{amsmath,amssymb,amsfonts}
\usepackage{algorithmic}
\usepackage{graphicx}
\usepackage{textcomp}
\usepackage{xcolor}
\usepackage{etoolbox}
\makeatletter
\patchcmd{\@makecaption}
  {\scshape}
  {}
  {}
 {}
\makeatother

\def\BibTeX{{\rm B\kern-.05em{\sc i\kern-.025em b}\kern-.08em
    T\kern-.1667em\lower.7ex\hbox{E}\kern-.125emX}}
\colorlet{shadecolor}{yellow}
\graphicspath{{../pdf/}{../jpeg/}}
\DeclareGraphicsExtensions{.pdf,.jpeg,.png}

\usepackage{array}
\usepackage{mdwmath}
\usepackage{mdwtab}
\usepackage{eqparbox}
\usepackage{url}
\usepackage{amsmath}
\let\labelindent\relax
\usepackage{enumitem}

\usepackage{tabularx,booktabs}
\newcolumntype{C}{>{\centering\arraybackslash}X} 
\usepackage{lipsum} 

\makeatletter
    \newcommand{\linebreakand}{%
      \end{@IEEEauthorhalign}
      \hfill\mbox{}\par
      \mbox{}\hfill\begin{@IEEEauthorhalign}
    }
    \makeatother
    
\begin{document}

    \title{Alzheimer's Disease Detection from Spontaneous Speech and Text: A review\\
    }

\author{\IEEEauthorblockN{
Vrindha M. K.\IEEEauthorrefmark{2} , Geethu V.\IEEEauthorrefmark{2} ,
Anurenjan P. R.\IEEEauthorrefmark{2} , Deepak S.\IEEEauthorrefmark{2} ,
Sreeni K. G.\IEEEauthorrefmark{3} }
\IEEEauthorblockA{\IEEEauthorrefmark{2}Computer Vision Lab, College of Engineering Trivandrum, Kerala.\\
\IEEEauthorrefmark{3}Rajiv Gandhi Institute of Technology, Kottayam, Kerala.\\
\IEEEauthorrefmark{2}\IEEEauthorrefmark{3}Affiliated to APJ Abdul Kalam Technological University, Trivandrum, Kerala, India.\\
Email: tve21ecsp17@cet.ac.in, tve21ecsp07@cet.ac.in, anurenjan@cet.ac.in, deepaks@cet.ac.in, sreenikg@rit.ac.in
}}

   


\maketitle

\begin{abstract}

 In the past decade, there has been a surge in research examining the use of voice and speech analysis as a means of detecting neurodegenerative diseases such as Alzheimer's. Many studies have shown that certain acoustic features can be used to differentiate between normal aging and Alzheimer's disease, and speech analysis has been found to be a cost-effective method of detecting Alzheimer's dementia. The aim of this review is to analyze the various algorithms used in speech-based detection and classification of Alzheimer's disease. A literature survey was conducted using databases such as Web of Science, Google Scholar, and Science Direct, and articles published from January 2020 to the present were included based on keywords such as ``Alzheimer's detection'', "speech," and "natural language processing." The ADReSS, Pitt corpus, and CCC datasets are commonly used for the analysis of dementia from speech, and this review focuses on the various acoustic and linguistic feature engineering-based classification models drawn from 15 studies.


Based on the findings of this study, it appears that a more accurate model for classifying Alzheimer's disease can be developed by considering both linguistic and acoustic data. The review suggests that speech signals can be a useful tool for detecting dementia and may serve as a reliable biomarker for efficiently identifying Alzheimer's disease.
\end{abstract}

\vspace{0.15cm}


\section{Introduction}

Alzheimer's Dementia (AD) is a progressive neurodegenerative disorder that is
characterized by the loss of subcortical neurons and synapses that begins in areas such as the
hippocampus and the entorhinal cortex \cite{terry1991physical}. In addition to neuronal and synaptic loss, more associative
areas begin to exhibit amyloid deposition and neurofibrillary tangles over time. As it spreads, patients develop additional cognitive and functional deficits
in domains such as attention, executive function, memory, and language \cite{nestor2004advances}. As the disease progresses, an Alzheimer's patient will experience severe memory loss and lose the ability to perform everyday activities.

In the past few decades, major progress has been made in the
development of biomarkers for AD diagnosis, such as $A\beta$ and
phosphorylated \cite{lee2019diagnosis}, neuroimaging techniques,
and neuropsychological tests. But these techniques are costlier and hence arises the need for cost-effective method for dementia detection. Distinguishing Alzheimer's from cognitive impairment due to aging is a challenging task. Many researches are being carried out in the natural language processing field to classify Alzheimer's disease using participants' speech. People suffering from dementia tend to show variations from normal people while describing an event. These variations and the related acoustic and linguistic features can be made in use for dementia classification. Figure 1 shows the word frequencies for the
healthy control and AD groups respectively \cite{yuan2021pauses}.

\begin{figure}[h]
\centering
\includegraphics[width=8cm]{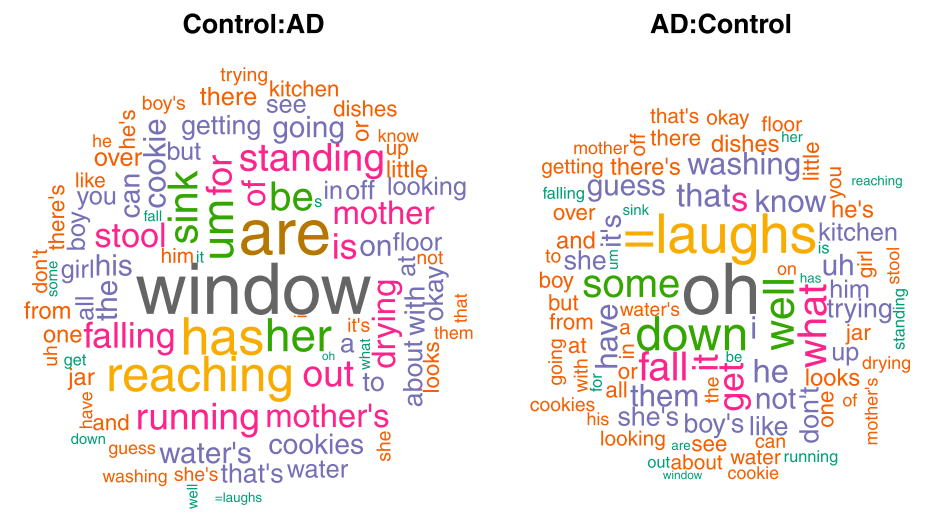}
\caption{The word cloud on the left highlights words that are more common among control subjects than AD; the word cloud on the right highlights words that
are more common among AD than control {\cite{yuan2021pauses}}}
\label{emotionmodel}
\end{figure}

 This paper is a systematic review on various acoustic and linguistic features for AD detection
using different classifiers. The feature engineering model studied mainly includes acoustic features such as Computational Paralinguistics Challenge features
set (ComParE) \cite{eyben2013recent},
X-vector, extended Geneva Minimalistic Acoustic Parameter Set (eGeMAPS) and Spectrogram features, and linguistic features such as Bidirectional Encoder Representations from Transformers (BERT) embeddings, Part-of-Speech (PoS), Perplexity etc,. The classifiers considered includes Artificial Neural Networks, Linear Discriminant Analysis, Decision Trees, K-Nearest Neighbour, Logistic Regression, Random Forest, and Support Vector Machine.

In this study, we reviewed various acoustic and linguistic feature engineering models that have been considered in the last two years, which commonly used ADReSS, Pitt corpus and CCC datasets.

The next section gives an overview of datasets, various feature extraction models and different classification methods.
\section{MATERIALS AND METHODS}
\label{sec3}
\subsection {Datasets}
Speech and language are strong biomarkers for AD detection. The researches in this field are mainly carried out using three different datasets, ADReSS challenge dataset \cite{luz2020alzheimer}, Dementia bank's Pitt corpus \cite{becker1994natural} and CCC. ADReSS challenge dataset is a subset of Pitt corpus that contains speech recordings describing the Cookie Theft
picture from the Boston Diagnostic Aphasia Examination. Contrastly to Demetia Bank's pitt corpus, the ADReSS challenge dataset is carefully matched to age and gender. This avoids bias in predication task. The details of the datasets are as shown in Table \ref{my-label1}.

\begin{table*}
\caption{Basic characteristics of patients in datasets}
\label{my-label1}
\centering
  \resizebox{0.5\textwidth}{!}{
\begin{tabular}{c cccc}
\toprule
Dataset & & & \multicolumn{2}{c}{{Class}} \\
& & & AD & Non-AD \\
\midrule
ADReSS & Train & Male & 24 & 24 \\
& & Female & 30 & 30 \\
ADReSS & Test & Male & 11 & 13 \\
& & Female & 11 & 13 \\
Pitt corpus & - & Male & 125 & 83 \\
& - & Female & 197 & 146 \\
\bottomrule
\end{tabular}
}
\end{table*}

\subsection{Preprocessing}

The speech recordings contains the interventions of interviewer, which needs to be removed and only the participant's clip need to be processed. The dataset also contains the transcriptions of each recording which are generated by automatic speech recognition (ASR) systems. Only the plain text in the transcripts need to be considered. Annotations of complex
events are removed.

\subsection{Feature Extraction}
Alzheimer's disease causes difficulties in verbal fluency, word finding, and word retrieval skills in addition to temporal alterations in speech \cite{szatloczki2015speaking}, hence they are strong biomarkers for AD detection. Various acoustic and linguistic features employed in different classification models are described below:

\begin{enumerate}[label=\roman*., itemsep=0pt, topsep=0pt]
\item \textbf{Acoustic features}:
The mainly used acoustic features includes ComParE, eGeMAPS, Mel Spectrogram, and  x-vectors. S. Haider et al. \cite{haider2019assessment} and Matej Martinc et al. \cite{martinc2021temporal} used eGeMAPS as the acoustic feature. Another acoustic feature, ComParE was considered by S. Haider et al. \cite{haider2019assessment}, Li et al. \cite{li2021comparative}, Zehra Shah et al. \cite{shah2021learning}, and Ablimit et al. \cite{ablimit2022exploring}. Generic acoustic emotion descriptors and their statistical functionals, such as temporal features and voicing-related low-level descriptors (LLDs), make up the ComParE feature set. The ComParE feature sets are extracted by
OpenSMILE \cite{eyben2010opensmile}. Mel spectrogram derived from the speech signals are very useful in providing accurate predictions. Davuluri et al. \cite{davuluri2022pre} and Flavio Bertini et al. \cite{bertini2022automatic} used Mel spectrogram as input feature to their classification model. Li et al. \cite{li2021comparative} and Ablimit et al. \cite{ablimit2022exploring} adopted representations based
on transfer learning like X-vector.

\hspace{0.6cm}

\item \textbf{Linguistic features}:
BERT is a kind of vector representation for text which has achieved state-of-the-art results in various language processing tasks. Li et al. \cite{li2021comparative} and Haulcy et al. \cite{haulcy2021classifying} have considered BERT embeddings. Other mainly used linguistic feature includes Linguistic Inquire and Word Count (LIWC), Part-of-Speech (PoS), and Perplexity which was adopted by Ablimit et al. \cite{ablimit2022exploring}. LIWC has been applied to mark individual, PoS tags words with similar grammatical properties into the same PoS tag
differences in cognitive processing, and Perplexity indicates how simple or difficult it is to predict the words in a text is indicated by the perplexity.
\end{enumerate}

\subsection{Classification Models}
In this subsection we looked into mostly used classifiers for AD detection.
\vspace{0.4 cm}
\subsubsection{Artificial Neural Network (ANN)}
A computational model known as an ANN is made up of a number of processing elements that, depending on their predefined activation functions, receive inputs and produce outputs. It includes Convolutional Neural Network (CNN), Deep Neural Network (DNN), Recurrent Neural Network (RNN), and Multi Layer Perceptron (MLP). LSTM is a type of RNN except that it has memorizing capability. They follows a supervised learning approach based on a
theory of association between cognitive elements. Among the reviewed articles \cite{meghanani2021exploration} employed a combination of CNN and LSTM \cite{meghanani2021recognition} used CNN based classification. \cite{liu2021detecting} adopted a combination of CNN, RNN and attention pooling which resulted in an accuracy of 82.59\%.

\vspace{0.4 cm}
\subsubsection{Linear Discriminant Analysis (LDA)}
 LDA, also known as normal discriminant analysis or discriminant function analysis, is a dimensionality reduction technique that is frequently utilized for supervised classification problems. It divides groups into two or more classes to show how different they are from one another. It is an unsupervised learning technique where the objective is
to maximize the relationship between the variance between groups and
the variance within the same group \cite{stanimirova2013robust}. It has been used as a classifier in various articles including \cite{haider2019assessment}, \cite{li2021comparative}, \cite{ablimit2022exploring}, and \cite{haulcy2021classifying}.
 \vspace{0.4 cm}
\subsubsection{K-Nearest Neighbour (KNN)}
 The K-NN algorithm stores all of the data that is available and uses similarity to classify a new data point. Because it is based on the idea that similar points can be seen close to one another, it is frequently used as a classification algorithm, even though it can be used for regression problems. Among the reviewed papers S. Haider et al. \cite{haider2019assessment} and Haulcy et al. \cite{haulcy2021classifying} have used KNN classifier as their classification model along with other models.
 \vspace{0.4 cm}
\subsubsection{Support Vector Machines (SVM)}
 It consists of building the hyperplane with maximum margin capable of
optimally separating two classes of a data set \cite{yahyaoui2018advances}. SVM classifier have been employed in many algorithms and it has achieved better classification accuracy when compared to other classifiers. \cite{haider2019assessment}, \cite{ammar2020language}, \cite{li2021comparative}, \cite{nasreen2021alzheimer}, \cite{shah2021learning}, \cite{ablimit2022exploring} and \cite{perez2022interpreting} used this classifier. Nasreen et al. \cite{nasreen2021alzheimer} achieved an accuracy of  90\% and Randa Ben Ammar et al. \cite{ammar2020language} achieved an accuracy of  91\%.

\vspace{0.4 cm}

 \subsubsection{Decision Trees (DT)}
 They are methods of supervised learning that are mostly used for classification issues. It is a classifier with a tree structure, with internal nodes representing a dataset's features, branches representing the decision rules, and each leaf node representing the result. The instances of the training set are
classified following the path from the root to a leaf, according to the result
of the tests along the path \cite{di2017exploring}. It has been used by Haulcy et al. \cite{haulcy2021classifying} and S. Haider et al. \cite{haider2019assessment} for classification.
\vspace{0.4 cm}
 \subsubsection{Logistic Regression (LR)}
 It is a supervised learning algorithm. It predicts the probability of target variables and is widely used in binary classification problems. Zehra Shah et al. \cite{shah2021learning} used this classifier and obtained an accuracy of 85.4\%.
 \vspace{0.2 cm}
 
 \subsubsection{Random Forest (RF)}
This has been employed by S. Haider et al. \cite{haider2019assessment}, Matej Martinc et al. \cite{martinc2021temporal} (accuracy 93.75\%) and Haulcy et al. \cite{haulcy2021classifying} (accuracy 85.4\%).

\begin{table*}
\caption{Recent Developments Alzheimer's detection using Speech and Text.}

\label{my-label}
\begin{tabularx}{\textwidth}{@{} l *{7}{C} c @{}}
\toprule
SL.No.  & Authors & Datasets & \multicolumn{2}{c}{{Features Extracted}}  & Classification Method & Results \\ &  & &Acoustic feature set& Linguistic feature set\\
\midrule

1 & S. Haider et al. \cite{haider2019assessment}      & Pitt Corpus & emobase, ComParE, eGeMAPS, MRCG functionals  & -  &  LDA, DT, KNN(with k=1), SVM and RF & Accuracy:  78.7\%    \\ 

2        & Liu et al. \cite{liu2021detecting}      & ADReSS & MFCC  & -  &  Combination of RNN, CNN and Attention pooling & Accuracy:  82.59\%, Precision: 85.29\%, Recall: 81.46\%, F1 score: 82.94\%     \\

3       & Davuluri et al. \cite{davuluri2022pre}       & ADReSS  & Mel Spectrogram & -& VGG-16 &  Accuracy: 95.83\%    \\ 
4      &Flavio Bertini et al. \cite{bertini2022automatic}      & Pitt Corpus & Mel Spectrogram & -  &  Auto encoder and MLP with data augmentation & Accuracy:  93.30\%, Precision: 90.7\%, Recall: 86.5\%, F1 score: 88.5\%    \\

5    &Randa Ben Ammar et al. \cite{ammar2020language}      & DementiaBank (TalkBank) &  - &  Temporal \& verbal disfluencies, Lexico-syntactic diversity, Word \& Utterance rate and MMSE score & SVM & Accuracy:  91\%,  Precision: 91\% \\

\addlinespace

6    & Meghanani et al. \cite{meghanani2021recognition}      & ADReSS & - & Bigrams, Trigrams, 4-grams, 5-grams  &  fastText, CNN & Accuracy:  83.33\%  for fastText model with bigrams \& trigrams    \\
7     & Li et al. \cite{li2021comparative}      & ADReSS & ComParE, X-vector  & TF-IDF vector, Linguistic feature sets, BERT embeddings  &  LDA, SVM and AT-LSTM & Accuracy:  67\% from X-vector by AT-LSTM \& 88\% on BERT by SVM    \\

8       & Nasreen et al. \cite{nasreen2021alzheimer}      & CCC & Interactional features  & Disfluency features  &  SVM, LR and MLP & Accuracy:  90\%, Recall: 90\%, Precision: 90\%, AUC: 89\%, and F1 score: 90\% on SVM classifier for combination of both features \\

9       &Matej Martinc et al. \cite{martinc2021temporal}      & ADReSS & eGeMAPS  &  GloVe embeddings& Combination of K-means clustering (with k=30) \& RF & Accuracy:  93.75\%, \\
10       & Zehra Shah et al. \cite{shah2021learning}      & ADReSS & AVEC 2013, ComParE 2013, emo-large, Jitter Shimmer, MFCC 1-16  & N-gram features, Type-Token Ratio (TTR), Mean Length of Utternace (MLU)  &  SVM \& LR for language features, Majority vote for combination of language and acoustic feature & Accuracy:  85.4\% for language feature using Logistic Regression \\

11      &Meghanani et al. \cite{meghanani2021exploration}      & ADReSS & log mel spectrogram, MFCC   &  -& CNN-LSTM, ResNet 18-LSTM, pBLSTM-CNN & Accuracy:  64.58\% by MFCC using CNN-LSTM \& 62.5\% by log mel spectrogram using ResNet 18-LSTM \\

12      &Ablimit et al. \cite{ablimit2022exploring}      & ADReSS, ILSE interviews &voice activity detection (VAD), ComParE, i-vectors, ECAPA-TDNN  & LIWC, PoS, Perplexity, PoS Perplexity& GMM, LDA, SVM &  UAR: 66.7\% \& 86\% on acoustic features   and 77.1\%  \& 83.8\% on linguistic features using SVM for ADReSS and ILSE respectively\\

13     &Haulcy et al. \cite{haulcy2021classifying}      & ADReSS & i-vectors, x-vectors & word vectors, BERT embeddings, LIWC features, CLAN features & SVM, RF, LDA, DT AND KNN (WITH K=1) & Accuracy:  85.4\% when trained on BERT embeddings using SVM \& RF\\

14      &Perez et al. \cite{perez2022interpreting}      & Pitt corpus & VAD duration features, pleasure arousal dominance (PAD), Harmonicity to Noise ratio & Phonemic features & eXtreme Gradient Boosting, ForestNet & UAR:  79\% using ForestNet for combination of duration, Phonemic \& Acoustic features\\

15      &Fritsch et al. \cite{fritsch2019automatic}      & Pitt Corpus & -  &  Perplexity& NNLM-LSTM & Accuracy:  85.6\%\\

\bottomrule
\end{tabularx}
\end{table*}


\section{Commonly used Evaluation Parameters}
Classification problems in machine learning field are frequently validated using accuracy, precision, recall, F1-score, and AUC.
\begin{enumerate} 
 \setlength{\itemsep}{-2ex}  
 \setlength{\parskip}{0ex} 
 \setlength{\parsep}{0ex}
\item
Accuracy (Acc):\\It is the most common evaluation indicator in classification problems, which is obtained as the ratio of number of correct samples predicted to the total number of samples.\hfil\\
\begin{equation}
    Acc= \frac{T_P+T_N}{T_P+T_N+F_P+F_N}  \times 100
\end{equation}
where, $T_P$, $T_N$, $F_P$, and $F_N$ represents the true positive, true negative, false positive, and false negative rates.

\hfil\break
\item
Precision ($P_r$):\\
Precision is ratio of true positive samples to the total number of samples.
\begin{equation}
     Precision= \frac{T_P}{T_P+F_P}  \times 100
\end{equation}

\hfil\break
\item
Recall ($R_e$): \\ The equation for calculating recall is specified in Equation (3). It is the proportion of true positive to the sum of true postives and false negatives.
\begin{equation}
     R_e= \frac{T_P}{T_P+F_N}  \times 100
\end{equation}

\hfil\break



\item
F1-score:\\It is the accuracy that combines precision and recall. A good F1 score indicates low false positives and low false negatives.
\begin{equation}\
 F-score= 2\times \frac{P_r \times R_e}{P_r + R_e} \times100
\end{equation}
 \hfil\break
 
 \item
AUC: \\ The AUC measures how well a model can distinguish between classes by counting the correct positive predictions  in comparison to the incorrect positive predictions at various thresholds.

\hfil\break
\end{enumerate}

\section{Conclusion}
Due to the fact that people with neuro-degenerative diseases frequently experience speech difficulties, research has been done on the use of numerous elements from spoken description to distinguish between a healthy person and a person suffering from cognitive decline. 

In this review we mainly focused on articles since 2020 that uses different acoustic and linguistic feature engineering based classification models for the detection of AD. We found that using linguistic features or both linguistic and acoustic features, produces better outcomes than relying just on acoustic features. It is clear that several algorithms make extensive use of the ComParE feature set and BERT embeddings. Mel spectrogram based feature extraction also achieves good results.

The reviewed articles have also proved that while considering the linguistic features from textual transcriptions of the speech signals, even though ASR transcripts attain comparable results, manual transcripts provides more accuracy. Different classifiers and many combinations of them have been in various classification algorithms. On comparing the different models it was clear that SVM is the mostly used classifier as it achieves better performance in AD classification.

Accuracy, precision, recall, and F1-score are the widely used evaluation metrics by researchers to assess the performance of classification models. This review comes to the conclusion that speech signal is a valuable tool for detecting dementia and can be utilised as a good biomarker for efficiently detecting AD.

\bibliographystyle{IEEEtran}
\bibliography{IEEEabrv,Bibliography}

\vfill

\end{document}